# Empirical Recurrence Rates for Seismic Signals on Planetary Surfaces

Ralph D. Lorenz[1*], Mark Panning[2, 3]

[1] Johns Hopkins Applied Physics Laboratory, Laurel, MD 20723, USA.

[2] Dept. of Geological Sciences, University of Florida, Gainesville, FL 32605, USA

[3] Now at Jet Propulsion Laboratory, California Institute of Technology, Pasadena, CA 91109, USA

* Corresponding Author:  tel: +1 443 778 2903  fax: +1 443 778 8939  email: (Ralph.lorenz@jhuapl.edu)

**Abstract**

We review the recurrence intervals as a function of ground motion amplitude at several terrestrial locations, and make the first interplanetary comparison with measurements on the Moon, Mars, Venus and Titan. This empirical approach gives an intuitive guide to the relative seismicity of these locations, without invoking interior models and specific sources: for example a Venera-14 observation of possible ground motion indicates a microseismic environment mid-way between noisy and quiet terrestrial locations; quiet terrestrial regions see a peak velocity amplitude in mm/s roughly equal to $0.4*N^{(-0.7)}$, where N is the number of events observed per year. The Apollo data show signals for a given recurrence rate are typically about 10,000 times smaller in amplitude than a quiet site on Earth, while Viking data masked for low-wind periods appears comparable with a quiet terrestrial site. Recurrence rate plots from in-situ measurements provide a convenient guide to expectations for seismic instrumentation on future planetary missions : while small geophones can discriminate terrestrial activity rates, observations with guidance accelerometers are typically too insensitive to provide meaningful constraints unless operated for long periods.

# 1. Introduction

Much of what is known about the Earth's interior is due to seismological studies and there is accordingly interest in using similar methods on other planetary bodies. A challenge often confronting mission designers (e.g. Lorenz, 2011) is an expectation of how much activity there might be, since this may determine the sensitivity required and/or the mission duration needed to observe some number of events.

One approach is to forward-model the problem: positing some annual average moment release through a distribution of event sizes, with events randomly located, and then to apply some estimate of crustal attenuation to derive the ground motion at a measurement station.

A simpler alternative, yet one that to our knowledge has not been systematically applied in the planetary literature, is to use data from a single seismic station as an analog for a future instrument.

# 2. Seismic Data

## 2.1 Earth

Our principal basis for comparison is a set of seismic records drawn from three Global Seismic Network stations. We analyzed 2 years of data (arbitrarily chosen to be 2010 and 2011). Data were downloaded from the Incoporated Research Institutions for Seismology Data Management Center (IRIS DMC) and processed using the seismological Python program ObsPy (Krischer et al., 2015). For each station, the processing was performed on the vertical BHZ channel, which included broadband data sampled at 20 Hz.

The three stations chosen are

1. BFO (Black Forest Observatory, Germany). BFO is a quiet station in the middle of a continent far from tectonic boundaries. This defines a quiet terrestrial case, where much of the ground motion spectrum is from distant events.

2. MIDW (Midway Island). This station is also in the middle of a plate (in this case an oceanic one), far from tectonic boundaries on a small coral atoll. However, the station has a higher ambient noise level due to the production of microseisms by nearby ocean waves, and Midway is also somewhat exposed to typhoons.

3. MSVF (Monasavu, Fiji). MSVF is also on an oceanic island, although the station is located further from shore than MIDW in a mountainous region in the island's interior. It is also near one of the most tectonically active regions in the world, the Fiji-Tonga subduction zone.

The seismic data were read in, and the instrument response was removed to produce ground velocity. The data were filtered between 0.01 and 2 Hz, and the peak amplitude in half-hour blocks was recorded (thus 17520 amplitudes in the two years). These peak amplitudes were converted to recurrence intervals by simply counting the number of blocks with peak amplitudes exceeding a given ground velocity and dividing by the record length. Throughout this paper we consider seismic signals with a nominal frequency of order 1 Hz: the large dynamic range of the parameter space discussed in this paper makes the conclusions robust to differences due to frequencies differing from 1 Hz by a factor of a few, and we neglect the modest differences in bandwidth associated with different sensors. Although displacement and acceleration also have virtues as metrics of ground motion, for the sake of using a single quantity, we display ground velocity. A useful guide to the presentation of seismic noise and data in its various forms is that by Bormann 1998; see also Lorenz (2011) and Bormann and Wielandt (2012).

The results are plotted in figure 1. The background noise amplitude is obvious in the recurrence intervals equal to about 100 events per year or more, and this clearly show how ocean noise dominates MIDW, while MSVF is generally quieter, and BFO quieter still. But the nearness of tectonic activity is more obvious in the strong but less frequent ground motion events, with MSVF reaching annual amplitudes of over 10 cm/s.

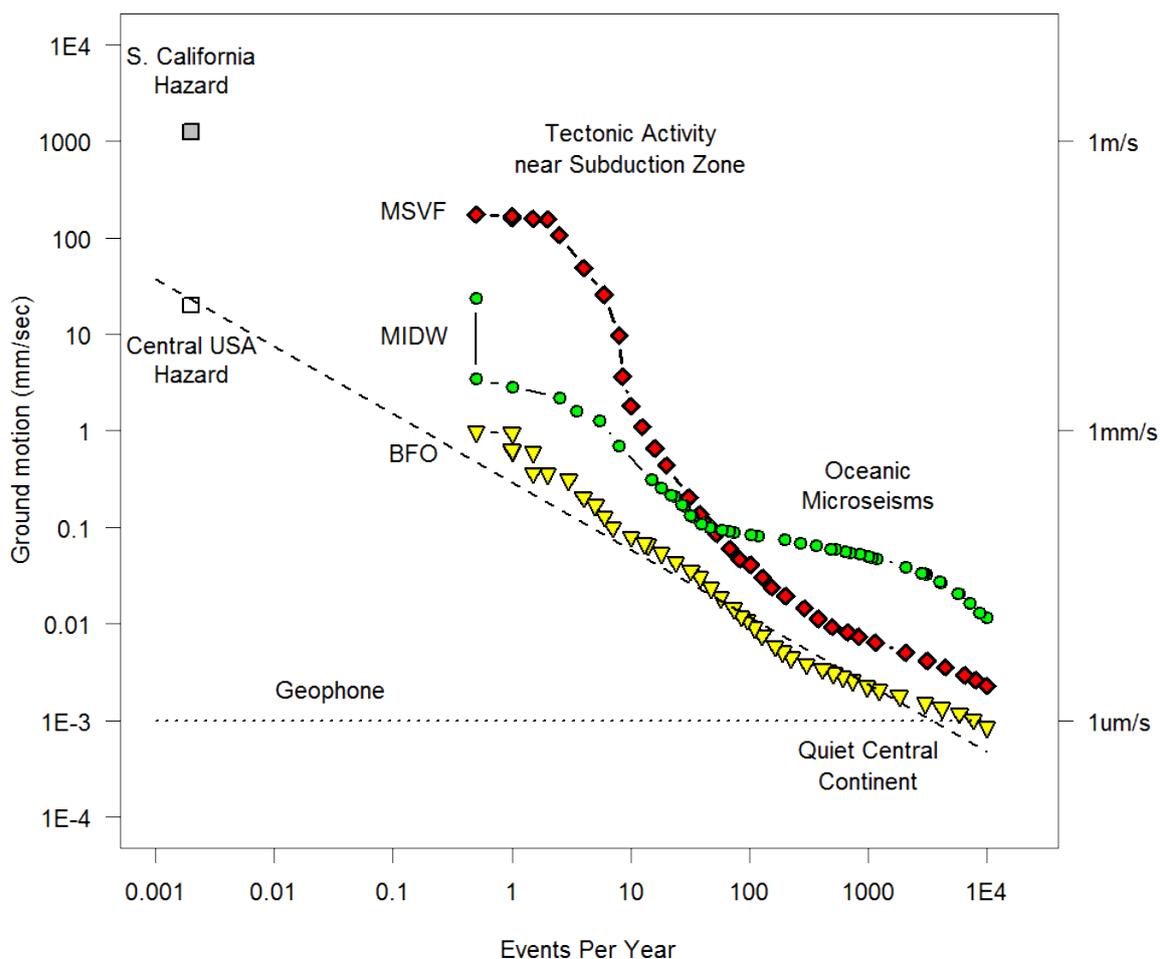

Figure 1. Peak ground motion amplitudes (in 30 minute windows over 2 years) for Monasavu (red diamonds), Midway (green circles) and Black Forest (yellow triangles). Squares show US long-term (~500 yr) seismic hazard. The dotted line is a notional 1 um/s measurement threshold, well within the capabilities of a small geophone. The dashed line is an empirical description of the quiet zone data, namely y=0.3*x^-0.7, with x and y units as defined on these plot axes.

Also shown in figure 1 are results derived from the US Geological Survey Seismic Hazard Map (Frankel et al., 2002) for very large but low probability ground motions. This product indicates the 1 Hz seismic acceleration for which a 10% probability of exceedance in 50 years is expected, for civil engineering applications (e.g. safety of dams, nuclear power plants etc.). In the middle of the continent (Texas to N. Dakota) this value is only about 0.015g, whereas the maximum, near the San Andreas fault in Southern California, is about 0.8g. We can interpret the acceleration as a velocity at 1 Hz by dividing by $2\pi$ to yield ~20 and 1200 mm/s respectively: a 10% chance of exceedance in 50 years translates into an expectation of one event in 500 years, or 0.002/yr. It is seen that the mid-continent point falls as a quite reasonable extrapolation of the BFO seismic recurrence intervals in our analysis above, and the California value is more consistent with the more active sites MIDW and MSVF, although the extrapolation of MSVF would likely plot significantly higher.

The dashed horizontal line in figure 1 indicates a 1 um/s threshold within the capabilities of a simple geophone (e.g. Rodgers, 1994 calculates a theoretical half-octave root-mean-square noise level of 1E-8 ms$^{-2}$ at 2Hz for a small (70g) L-22D geophone, corresponding to a ground motion noise of ~0.1 um/s). If an instrument with such a threshold were deployed and operated for 1 day, it would detect ~10 events even at a quiet site such as BFO, or some dozens of events at MSVF, and hundreds at MIDW, discriminating the different seismic environment at these sites. Note that this approach does not make any assumptions about the source of the ground motion, which may be tectonic events or ambient noise excited by the oceans. This, however, serves as a useful benchmark for planetary seismological characterization.

**2.2 Moon**

Long-lived highly sensitive seismometers (e.g. Latham et al., 1973; Nakamura et al., 1982) were emplaced on the Moon by Apollo 12, 14, 15 and 16 ; similar but short-lived instrumentation was deployed by Apollo 11.

Here we use the updated catalog ("levent.1008") of long-period seismometer events developed by Y. Nakamura (Nakamura et al., 1991). This catalog includes meteorite (and artificial) impact signals as well as interior seismic events. It may be noted that a significant proportion of the lunar seismic event population is tidally-modulated. We show the Apollo 12 record (the longest) in figure 2. The amplitudes listed in this catalog are listed as amplitudes in mm in a "compressed plot" (see the explanation in Nakamura et al., 1991): making the rather arbitrary assumption of a 1 Hz frequency, we multiply the data values in the catalog by 0.4 nm/s (determined from scaling the catalog amplitudes to the known ground velocity from the Saturn IV-B stage impacts) to plot them on figure 2. As an independent check on this scaling, we also compared the catalog amplitudes with the waveform amplitudes for the six largest recorded events of 1972 at the A12 station. When filtered between 0.05 and 0.5 Hz, these produced an average scaling factor of 0.45 +/- 0.19 nm/s, suggesting the 0.4 nm/s value is likely reasonable within a factor of 2.

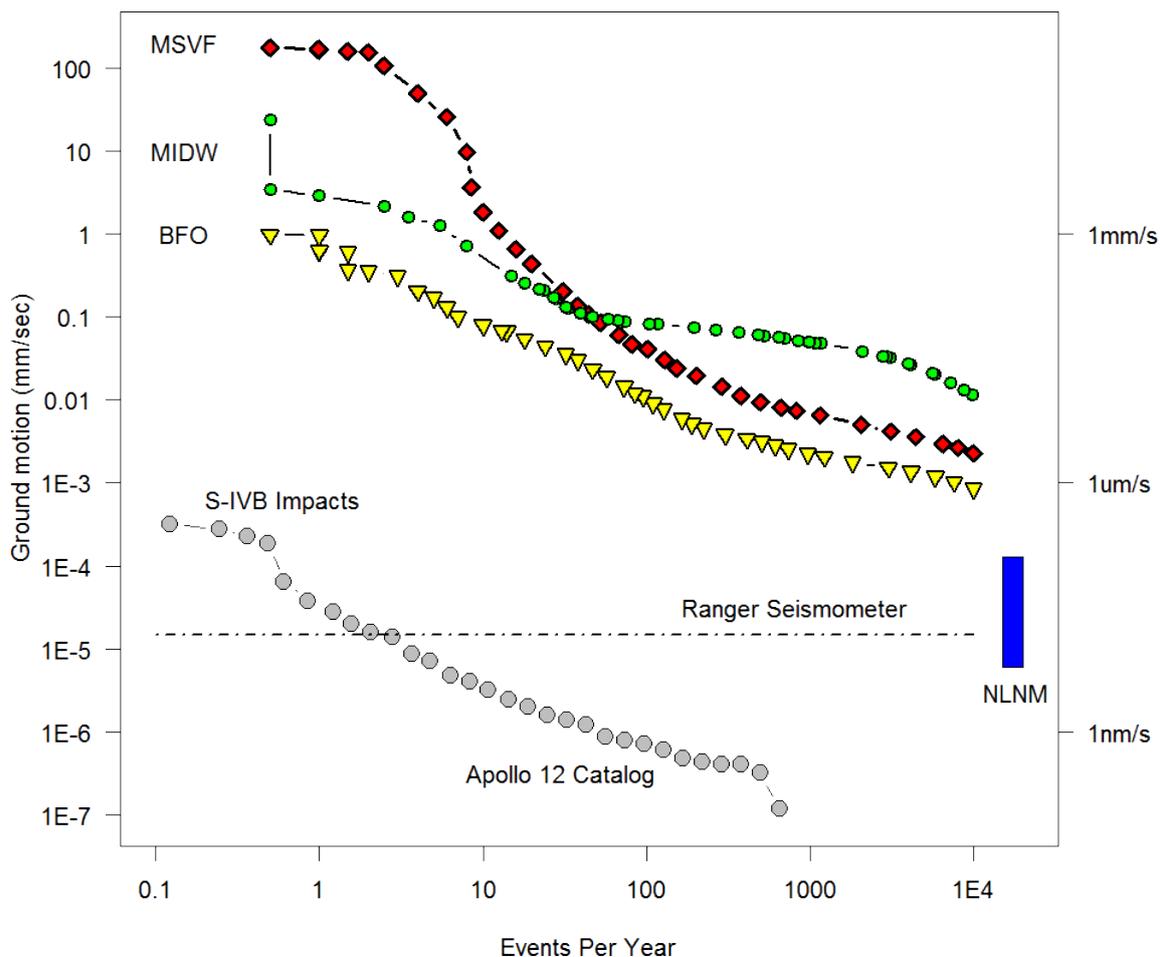

Figure 2. The Apollo 12 catalog, compared with the terrestrial sites in figure 1 – notice the expanded vertical scale to allow portrayal of the very small lunar seismic signals. Although for a given recurrence rate (e.g. 100/yr) the lunar activity is about 4 orders of magnitude smaller than the quiet continental site, the slope of the distribution is similar. Three of the four largest events recorded by the Apollo seismometer were artificial impacts of the Saturn IVB rocket stages on the moon. The dot-dash line shows the sensitivity of the Ranger seismometer. The blue box at right is the New global Low-Noise Model (NLNM) description of continuous terrestrial background at quiet locations by Peterson (1993).

The moon is very quiet seismically – while this is well-known, the recurrence-interval portrayal of lunar and terrestrial events in figure 2 makes this difference especially apparent. The size of a given rate event (e.g. 100 per year) appears to be ~10,000 times larger even in quiet areas on Earth; natural events of a given size – in fact the curves do not even really overlap – are observable about a million times less frequently at a station on the moon than on the Earth.

It is of interest that the very first seismic experiment launched to another planetary body, namely the seismometer on the Ranger 1 capsule, would likely not have observed anything. This instrument (Lehman et al., 1962), a single-axis coil-spring-magnet arrangement (i.e. a sensitive geophone) was impact-tolerant to 3000g and had a novel vented-liquid leveling system. Its output was modulated directly onto a carrier radio signal, and the sensitivity was expressed as a magnification of 1.7 million at its most sensitive frequency of 4Hz, to produce a detectable (1mm) trace on a chart recorder. The corresponding velocity sensitivity is therefore about 15 nm/s, shown as a dash-dot line in figure 2. At this threshold, the (much more sensitive) Apollo record suggests only a few events per year would be detected, but the battery capacity of the Ranger capsule provided only for ~30 days of operation, so most likely no events would have been recorded on any given mission.

**2.3 Mars**

The Viking landers had the principal goal of detecting life on the Martian surface, and geophysical and meteorological instrumentation suffered a number of compromises in accommodation (Anderson et al., 1972; 1976; 1977). Most notably, the seismometer was mounted on the lander deck. Although for seismic periods of interest the coupling of the instrument with the ground was adequate (contrary to what is often asserted), this arrangement did cause the noise floor of the instrument to be often rather high due to wind-induced motion of the lander body on its compliant legs. For extended periods (especially at night) when wind was low, the instrument was useful. The Viking landers each carried a sophisticated seismometer, and that of Viking 2 operated for over 500 sols. Although the data are usually dismissed as of limited value due to the susceptibility of the lander-mounted instrument to wind noise, they were able to exclude high global seismicity on Mars (see e.g. Goins, and Lazarewicz, 1979), with only one candidate local seismic event being identified (Anderson et al., 1977).

Here we use a summary data product recently delivered (Lorenz et al., in preparation) to the NASA Planetary Data System Geosciences node (dataset ID VL2-M-SEIS-5-RDR-V1.0). This summary product identifies the maximum amplitude recorded in each record (~51s long) of the most-extensively-used compressed 'event' mode: it also lists the nearest wind speed measurement and the time difference between two, allowing data to be selected for given known wind speed conditions.

The full record is shown in Figure 3. The Viking instrument was generally operated in its highest-gain state, with each data number corresponding to roughly 50 nm/s, but the data were recorded with only 7 bits of digital resolution (i.e. 1 to 127 units) thus the dynamic range is somewhat limited and the curve flattens towards the left, probably indicating saturation.

Note that the Viking record includes lander-generated noise, from mechanical operations including small signals from the tape recorder and camera and much larger vibrations from movements of the high-gain antenna and sampling arm, as well as vibrations from sample agitation in the X-ray fluorescence instrument. Since these disturbances were not fully documented, it is impossible to eliminate their contribution to the record: however, it should be noted that most of the event mode data were recorded in the middle of the night when other lander activities were minimized.

Further, as is well-known, the Viking seismometer was mounted on the lander deck, which was held above the ground on spring-loaded legs, making the system susceptible to wind noise (e.g. Anderson et al., 1977; Nakamura and Anderson, 1979) which was noticeable for winds above 3 m/s. Again, because the bulk of the data were recorded in the middle of the night, the winds were generally low.

Figure 3 shows the Viking data selected when the wind was known to be <3 m/s. These data have not been rescaled to normalize for the shorter observing time – thus the difference in the curves indicates when 'good' data are available with a given amplitude (although this curve is also contaminated with lander noise which has not been documented.) If the wind effect could be eliminated (e.g. by a wind shield as to be implemented on the forthcoming InSight mission) the resulting recurrence amplitudes would be between these two curves.

A single candidate seismic event was identified (Anderson et al., 1977) in the Viking record, on Sol 80, interpreted as a Magnitude 2.7 event about 60km away. The peak signal envelope was 40 DN indicating a ground motion of ~2 um/s. Given the total observing time of ~0.25 years, we indicate this putative observed seismicity rate as a 4x/yr rate of this amplitude.

Since Viking, it may be noted that the Mars Exploration Rover Opportunity (Arvidson et al., 2011) has conducted 'Marsquake' observations on a few Sols using the accelerometers in its Litton LN-200S inertial measurement unit (IMU) as crude seismometers. The manufacturer's specification (Northrop-Grumman , 2013) of the accelerometer noise at 1 Hz is notionally only 35 micro-g (i.e. 55 um/s in velocity terms) although the recording accuracy implemented on the vehicle may have been somewhat poorer. Since no positive detections have been reported in the four Sols indicated in Arvidson et al. (2011), we indicate an upper limit of 55 um/s at 0.011 events/yr on figure 3.

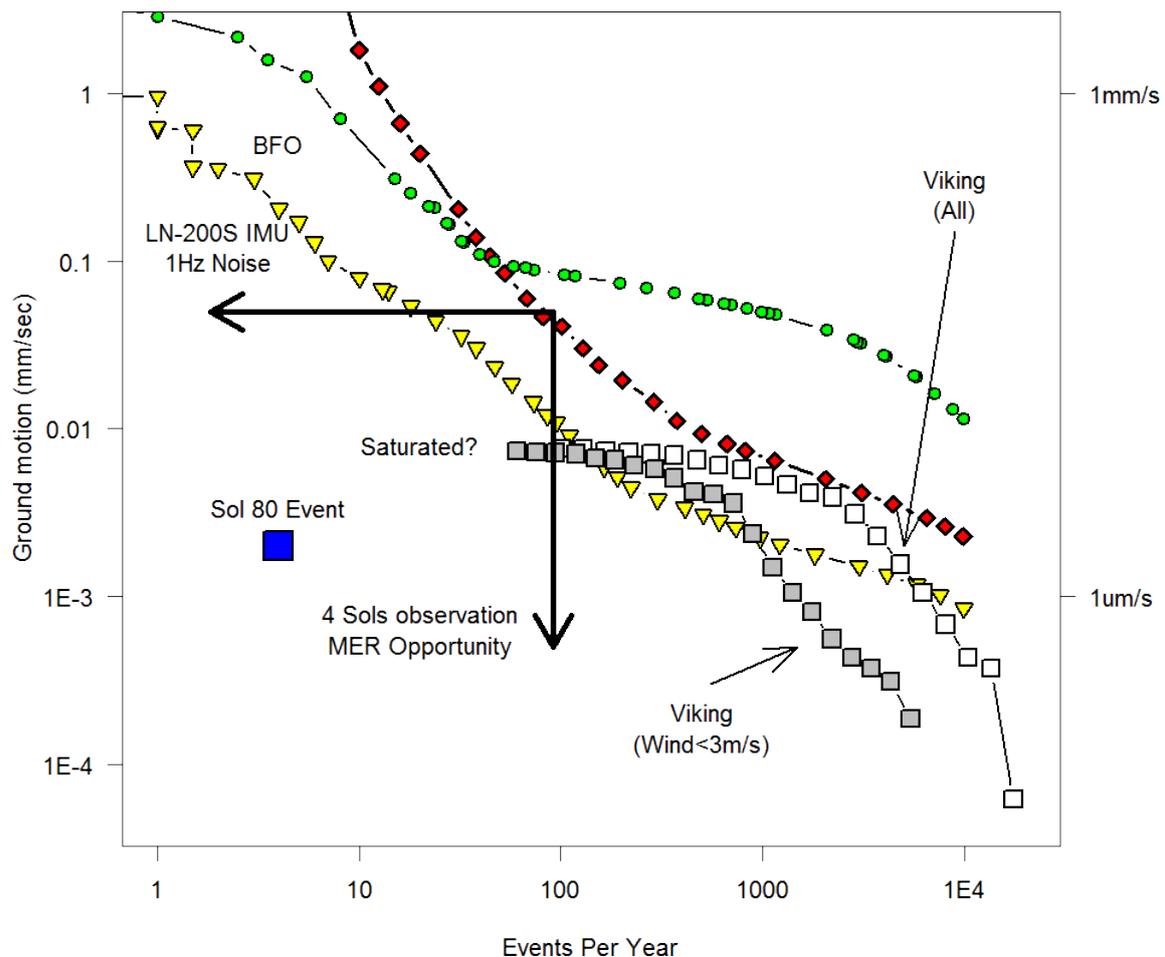

Figure 3. Mars seismometer record. Open squares indicate the full event mode record, spanning about 0.25 Earth years of observation. Grey squares indicate those records where the nearest wind measurement was less than 3 m/s, indicating no significant wind noise contribution. Both curves become flat below 100 events/yr due to instrument saturation. The large blue square indicates the observed seismicity implied by the tentative identification of a single 2um/s Marsquake on Sol 80. The rather weak constraint inferred from non-reporting of results from the Opportunity rover IMU observation is indicated by thick arrows.

**2.4 Venus**

Ksanfomaliti et al. (1982) reported the operation of geophone-type vertical seismometers on the Venera-13 and -14 landers. Although data acquisition was somewhat rudimentary (continuous recording not being possible, some raw signals were recorded and a threshold counter was implemented. Although Venera-13 indicated no seismic signals, Venera-14 may have recorded one displacement of 80E-6cm (i.e. 0.8 um) and one 'somewhat smaller' in the approximately 1-hr of surface operation. The frequency of these signals is not known – assuming (arbitrarily: the instrument resonance frequency was 24Hz) a value of 4Hz and (again arbitrarily) an amplitude of the $2^{nd}$ signal as half the first, we derive ground velocities of 3.2 and 1.6 um/s. These events are not obviously correlated with wind and so may be real ground motion.

Plotting these points (figure 4) on the recurrence rate diagram (at a rate of 1 and 2 events per landed operation time of 72 minutes) used for the other datasets allows us to critically examine the plausibility of this observation, which has not been widely discussed in the literature.

In fact, the two points define a cumulative recurrence function not unlike that at Fiji – intermediate between the quiet continental and noisy mid-ocean microseismic environments at BFO and MIDW. While Venus lacks oceans, it is not improbable that its massive atmosphere can couple meteorological variations into the ground (see Lorenz, 2011). Another possibility is of course that the Venera 14 site was nearer a site of volcanic or tectonic activity than Venera 13.

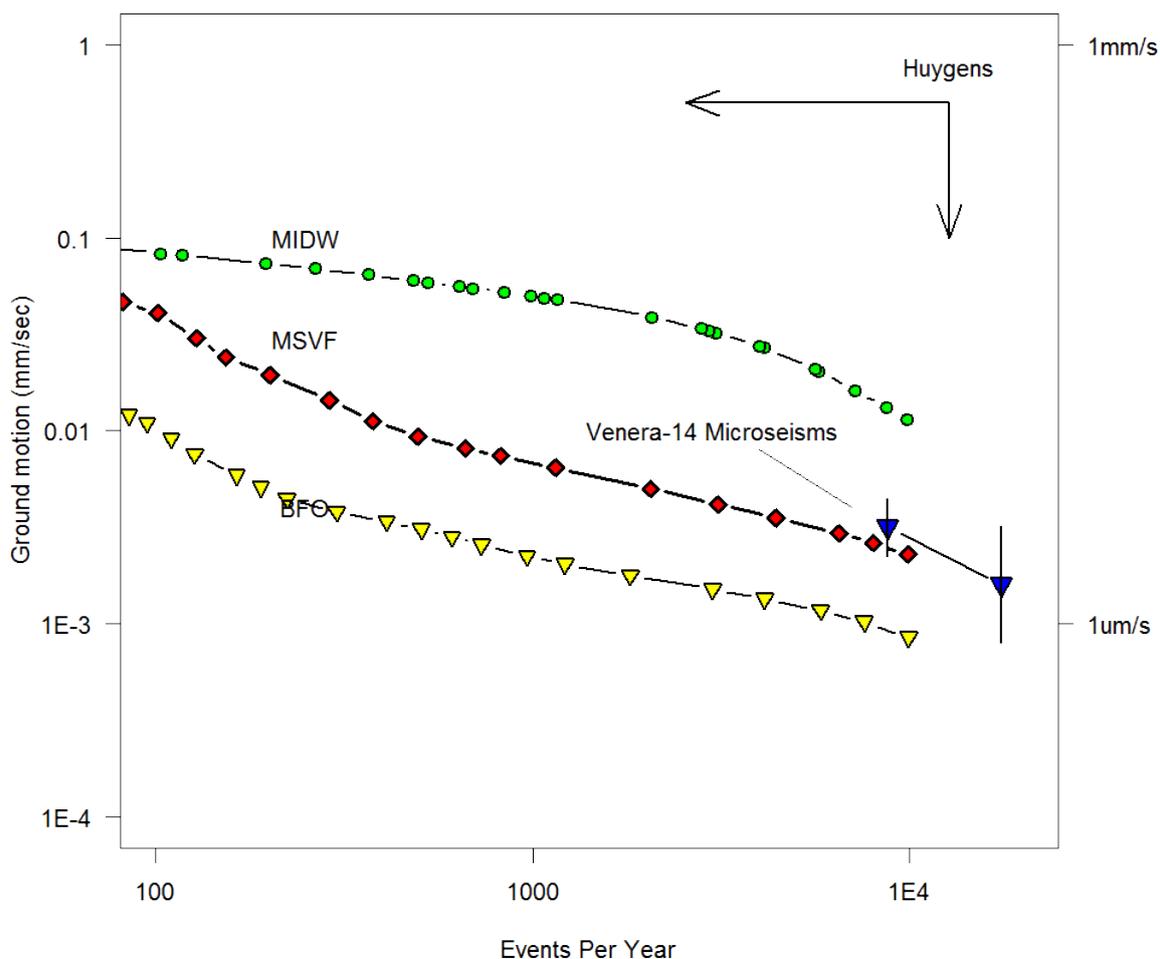

Figure 4. Venus and Titan seismic observations. The Venera-14 observation by Ksanfomaliti et al. (1982) plotted as blue triangles against the terrestrial recurrence curves appears not unlike an active terrestrial site (MSVF), but with less activity than a site that sees extensive microseismic activity due to ocean waves (Midway). The Huygens accelerometer measurement does not give a strong constraint on Titan seismicity (nor was it expected to).

**2.5 Titan**

Although the Huygens probe specification did not require post-landing operation, the impact was relatively benign, and the probe continued operation for several hours, of which 72 minutes were received before the Cassini spacecraft set below the horizon as seen from the probe. Of this period, the first ~32 minutes included measurements recorded at 1.75Hz by a sensitive accelerometer (Hathi et al., 2007).

Inspection of the record (Dataset ID "HASI_L3_ACCS_SERVO" in the Huygens archive at the PDS Atmospheres node) shows no obvious signal : two isolated spikes of about 0.006 ms$^{-2}$ are seen, a multiple of about 8 times the quantization interval. Since these are isolated single samples, it seems most likely these are digital noise rather than mechanical motion. We therefore assert for the half-hour observation interval that no seismic signals with a velocity amplitude of <0.5 mm/s were detected, a rather weak constraint, as seen in figure 4.

## 3. Discussion

By definition, all of these recurrence rate plots will be monotonically decreasing; however, there is significant variation in the slopes of the curves. For the terrestrial records (figure 1), all three show a distinctive break in slope somewhere between 10 and 1000 events per year, with the curve having a slope near 1 in log-log space to the left of the break, but significantly shallower to the right. This is consistent with the high recurrence rate "events" being primarily caused by oceanic microseisms, which create the background noise at all seismic stations even in quiet continental sites like BFO (e.g. Peterson, 1993), while the larger amplitudes are caused by tectonic events. These different sources have different characteristic size-frequency relationships.

Tectonic events have long been shown to follow a size-frequency relationship defined by the Gutenberg-Richter relationship (Gutenberg and Richter, 1944), which predicts a slope of -1 between log recurrence rates and earthquake magnitude (which scales with the log of seismic amplitude), while the microseismic noise from the ocean shows less temporal variation at a particular site, thus producing a flatter recurrence curve.

For the other planetary datasets, there may be a similar variation in slope as a function of the exciting force for the recorded events. The Apollo data shows a slope consistent with the tectonic events of the Earth data, which is reasonable since the source of the data was a curated moonquake catalog, and there is no available ocean or atmosphere to excite a background ambient noise. For the Viking data, and possibly the Venera data as well although there are only wo points and thus not a well-constrained slope, we see a much flatter curve, which suggests the source of the seismic data recorded in these situations is ambient noise caused by wind and pressure noise in the thin atmosphere of Mars and the dense atmosphere of Venus.

## 4. Conclusions

A recurrence rate plot offers an illuminating portrayal of the seismic character of different planetary surface environments. Such plots provide a useful guide to expectations for the required sensitivity and measurement duration for landed seismic studies. For example, to detect activity at some quiet areas on Earth, a guidance accelerometer (such as carried on Opportunity or Huygens) could discriminate between zero activity, quiet mid-continental and seismically-active areas on Earth, but only if operated for several weeks such that hundred-microns-per-second events with a recurrence rate of tens per year would statistically be encountered.  Even simple geophones offer rather higher sensitivity, and the Venera-14 report of Venus microseisms detected with such an instrument in only an hour of operation is consistent with 'noisy' environments on Earth.

**Acknowledgements**

RL acknowledges the partial support of NASA Mars Data Analysis Program  Grant  NNX12AJ47G. RL thanks Y. Nakamura and J. Murphy for assistance with the Viking data. MP acknowledges N. Schmerr for assistance with the Apollo data.